\documentclass[preprint,12pt,authoryear]{elsarticle}

\usepackage[]{graphicx}
\chardef\bslash=`\\ 

\newcommand{\bmX}{\mbox{\boldmath $X$}}
\newcommand{\bmM}{\mbox{\boldmath $M$}}

\newcommand{\bmZ}{\mbox{\boldmath $Z$}}

\newcommand{\bmU}{\mbox{\boldmath $U$}}

\newcommand{\bmzero}{\mbox{\bf 0}}
\newcommand{\bmbeta}{\mbox{\boldmath $\beta$}}

\newcommand{\bmtheta}{\mbox{\boldmath $\theta$}}

\newcommand{\ba}{\begin{array}}
\newcommand{\ea}{\end{array}}

\newcommand{\bea}{\begin{eqnarray*}}
\newcommand{\eea}{\end{eqnarray*}}

\newcommand{\beaa}{\begin{eqnarray}}
\newcommand{\eeaa}{\end{eqnarray}}

\newcommand{\be}{\begin{enumerate}}
\newcommand{\ee}{\end{enumerate}}

\usepackage{amssymb}

\journal{Computational Statistics and Data Analysis}

\begin{document}

\begin{frontmatter}



\title{On estimating causal controlled direct and mediator effects for count outcomes without assuming sequential ignorability}

 \author[label1]{Cheng Zheng}
  \author[label2]{David C. Atkins}
   \author[label2]{Melissa A. Lewis}
    \author[label3,label4]{Xiao-Hua Zhou\corref{cor1}}
    \cortext[cor1]{Corresponding author \ead{azhou@uw.edu}}
        
 \address[label1]{School of Public Health, University of Wisconsin-Milwaukee\\Milwaukee, WI 53205, USA}
 \address[label2]{Center for the Study of Health and Risk Behavior, University of Washington\\ Seattle, WA 98105, USA}
 \address[label3]{HSR\& D, VA Puget Sound Health Care System\\ Seattle, WA 98101, USA}
 \address[label4]{Department of Biostatistics, University of Washington\\ Seattle, WA 98195, USA}

\begin{abstract}
Causal mediation analysis is an important statistical method in social and medical studies, as it can provide insights about why an intervention works and inform the development of future interventions. Currently, most causal mediation methods focus on mediation effects defined on a mean scale. However, in health-risk studies, such as alcohol or risky sex, outcomes are typically count data and heavily skewed. Thus, mediation effects in these setting would be natural on a rate ratio scale, such as in Poisson and negative binomial regression methods. Existing methods also mainly rely on the assumption of no unmeasured confounding between mediator and outcome. To allow for potential confounders between the mediator and outcome, we define the direct and mediator effects on a new scale and propose a multiplicative structural mean model for mediation analysis with count outcomes. The estimator is compared with both Poisson and negative binomial regression methods assuming sequential ignorability using a simulation study and a real world example about an alcohol-related intervention study. Mediation analyses using the new methods confirm the study hypothesis that the intervention decreases drinking by decreasing individual's normative perceptions of alcohol use.

\end{abstract}


\end{frontmatter}


\section{Introduction}
\label{sec:In}
Mediation analyses play an important role in clinical research. For example, the general efficacy for a number of behavioral treatments has largely been established, and research is increasingly focused on treatment mechanisms -- how do interventions work \citep{Nock_2007}? Similarly, in the medical research, the question related to mediation analysis began to be raised recently. One example is how genetic variants affect lung cancer through changing cigarette smoking \citep{Valeri_2014}. Another example is to study how socioeconomic status affect number of decayed, filled or missing teeth (DMFT) through number of dentist visit \citep{Wang_2013}. Other example that is not called mediation analysis, but share the same mathematical formula are surrogate evaluation, where the effect of surrogate on the outcome can be considered as mediator effect, which will be rigorously defined in following sections. Modern interventions often have multiple components, with multiple hypothesized mechanisms of action. Thus, mediation analysis is needed to determine the active ingredients of interventions to improve the strength of interventions by focusing future intervention on these ingredients.

Until now, the most popular framework for performing mediation analyses in applied research is given by \citet{Baron_1986}, which has been extended by \citet{MacKinnon_2007}. The traditional approach describes a clear set of regression equations that depict the structural relationships among treatment, mediator, and outcome. Although these methods are extremely common in psychology and related fields, the causal interpretation of the result is not gauranteed. \citet{Imai_2010} showed that under sequential ignorability assumption, which implies that there is no unmeasured confounder between mediator and outcome, Baron and Kenny's approach has causal interpretation.
	
To establish the causal interpretation of mediation analysis, several frameworks have been developed within Rubin's causal model \citep{Rubin_1974}, including principle stratification models \citep{Gallop_2009} and counterfactual models, such as structural mean model \citep{Robins_1992}. Principal strata method is often used in the context of compliance of the intervention and often needs the level of mediator and exposure to be small, for example, binary. It often also needs specific distribution assumption for the residuals. Moreover, \citet{VanderWeele_2011} showed that the effects defined in the principal strata framework are different from that of the counterfactual model and traditional approach \citep{Baron_1986}. So in this paper, we will follow the counterfactual model, which produce estimators that can be easily compared with traditional regression approach.

As usual, we denote the intervention as $Z$ and the mediator as $M$, the potential outcome as $Y_i^{zm}$, which represents the potential outcome for subject $i$ if we manipulate its intervention at level $z$ and its mediators at level $m$. The controlled average direct and mediator effects are usually defined as follows:\\\\
Controlled Mediator Effect: $E(Y_i^{zm}-Y_i^{z0})$,\\
Controlled Direct Effect: $E(Y_i^{zm}-EY_i^{0m})$.\\ \\
The controlled mediator effect represent the causal effect of the mediator on the outcome and the controlled direct effect represent the effect of intervention on the outcome that is not through mediator. The controlled mediator effect provide us the information whether we should aim at changing certain mediator during future intervention development. Under so called ``sequential ignorability", also known as no unmeasured confounder assumption, estimators based on counterfactual models are shown to be the same as the one given by traditional regression approach \citep{Imai_2010a}. When sequential ignorability assumption is assumed, Imai's method \citep{Imai_2010} can be used to nonparametrically estimate the natural direct and indirect effects for both linear and nonlinear models. However, without randomization of mediator levels to individuals, this assumption cannot be guaranteed due to the potentially unidentified or unmeasured confounders between the mediator and the outcome.

For count regression, when using a Poisson or negative regression model, it is straightforward to consider the effect in terms of rate ratio \citep{Maldonade_2002, Frome_1985}. So by extending the traditional effects \citep{VanderWeele_2013} to rates ratio scale, we can define our effects of interest as follows:\\ \\
Controlled Mediator Effect: $\log(EY_i^{zm})-\log(EY_i^{z0})$,\\
Controlled Direct Effect: $\log(EY_i^{zm})-\log(EY_i^{0m})$.\\ \\
We can also use the risk difference (RD) as effect scale, but the effect is less likely to be homogeneous. Although different definitions might give different mediation effects, we know that if there is perfect mediation, then we should have $Z$ and $Y$ independent if we fix $M$, and thus indirect effect defined by any scale should be 0. In this sense, the different definitions are consistent.

Estimating effects related to mediation analysis is difficult when the sequential ignorability assumption does not hold. When the outcome is continuous and a linear model is used, a special structural mean model (SMM) called a rank preserving model (RPM) \citep{Tenhave_2007} has been proposed to relax the key, sequential ignorability assumption. However, the rank preserving assumption is implausible when outcome is a 'count' type variable. Previously, when considering compliance status as mediation and assuming exclusion criteria, several multiplicative structural mean model have been proposed \citep{Vansteelandt_2003, Robins_1994, Tan_2010}. However, the exclusion restriction assumption used in those methods is not appropriate in the case of mediation analysis and when interest focuses on not only the mediator effect but also the direct effect.

In this paper, we propose a new method using a multiplicative structural mean model that does not rely on either sequential ignorability assumption or exclusion restriction assumption for making valid inference on controlled indirect and mediator effects. We choose the log-linear model for the reason that it is commonly used in the analysis of count outcome data (Poisson regression and negative binomial regression). Our method can serve as an important alternative to analyze data where sequential ignorability assumption is questionable. Also, our method is useful when the targeted causal effect we are interested in is rates ratio rather than difference in mean.

We would like to point out that the effects estimated by our method and Imai's method \citep{Imai_2010} are different. We estimate controlled effects rather than the natural effects. A discussion of these two kinds of direct effect is reviewed by \citet{Petersen_2006}. Beyond that, we estimate the mediator effect instead of the indirect effect. The mediator effect is useful since it assesses how strong the causal relationship is between the mediator and the outcome. A strong effect suggests that we might improve future intervention by enhance the component that can change that mediator. When the linear model is used for the continuous outcome, we know that the indirect effect can be written as the product of the effect of the intervention on the mediator and mediator effect. Similarly, the existence of both mediator effect and intervention effect on the mediator suggest the existence of indirect effect for nonlinear models.

In Section~2, we introduce the motivating example and summarize previous analysis for the main effect. In Section~3, we give notation of our model, followed by estimation method and asymptotic theory. In Section~4, we give the simulation result. In Section~5, we applied our method to the real example and discuss the results. In Section~6, we give a further discussion of the model with certain extensions.

\section{Applied Example: Interventions for Problematic Alcohol Use and Alcohol-Related Risky Sexual Behavior}
\label{sec:ME}
The motivating example is a randomized controlled trial studying the effects of web-based personalized normative feedback (PNF) interventions targeting high-risk alcohol use and alcohol-related risky sexual behavior (RSB) among college students (Lewis et al., in press). Perceiving that peers drink more than they actually do (i.e., normative misperception) is associated with heavier drinking and experiencing negative consequences \citep{Borsari_2001}. PNF interventions \citep{Lewis_2006} were designed to correct overestimated normative perceptions and had been shown to reduce heavy drinking and alcohol-related problems \citep{Neighbors_2004}. In the example study, after screening for eligibility, 480 participants were randomized to four intervention groups: Alcohol only PNF (n=119), alcohol-related RSB only PNF (n=121), Control (n=121) and Combined alcohol and alcohol-related RSB PNF (n=119). Personalized normative feedback (PNF) interventions \citep{Lewis_2006} were designed to correct normative misperceptions (i.e., students often assume that typical drinking by college students was much higher than actual). Depending on condition, the different treatment groups received three pieces of information for alcohol, alcohol-related risky sex, or combined alcohol and alcohol-related risky sex: 1) personal risk behavior, 2) perceived peer risk behavior, and 3) actual peer risk behavior. Previous analysis (Lewis et al., in press) showed that groups with an alcohol component (i.e,. alcohol only PNF, combined alcohol and alcohol-related risky sex PNF) had significantly lower normative perceptions (i.e., mediator) and drinks per week, drinking per occasion, and drinking frequency at both 3 and 6 months. Though intervention conditions including alcohol-related RSB had lower normative perceptions on frequency of drinking prior to sex and drinks consumed prior to sex, the corresponding outcome decreased only significant for frequency of drinking prior to sex at month 3.

Each intervention condition targeted a particular type of normative perception by the participant, and thus, the normative perception after intervention is a hypothesized mediator of intervention efficacy. Theoretically, any changes in an outcome should be the result of changes in perceived norms. However, it is impossible to randomize levels of normative perception to participants, and hence, this hypothesis is open to measured and unmeasured confounders. A graphical representation for our hypothesized mechanism is shown in Figure 1. Note that here we allow the existence of unmeasured confounders (such as genotype, environmental variables) between the mediator and the outcome. All the outcomes and norms were recorded at both 3 and 6 months after the randomization.

\begin{figure}
\centering
\includegraphics[width=4in]{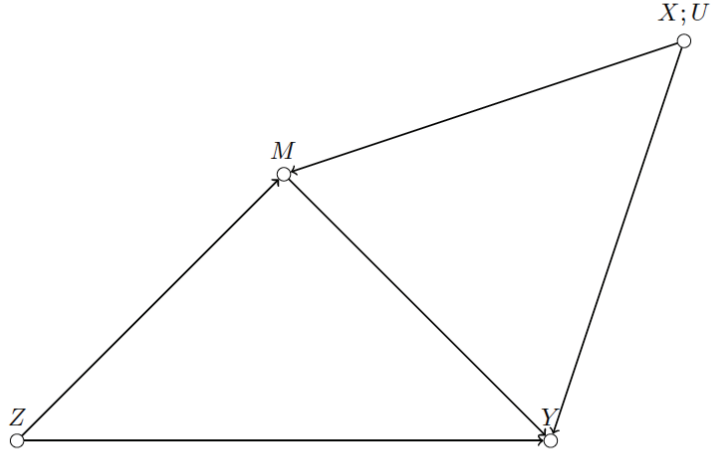}
\caption{Hypothesized mediation mechanism using directed acyclic graph. Here $Z$ is intervention, $M$ is mediator (i.e. perceptive norm), $Y$ is the outcome (i.e. drinking behavior), $X$ is observed baseline covariate, and $U$ is unmeasured confounder.}
\end{figure}

In this paper, we focus on drinks per occasion as the outcome. The natural mediators are the normative perception of drinks per occasion and drinking frequency. Primary analyses (Lewis et al., in press) suggested this outcome did not have excessive zero and the Poisson regression might be appropriate. In Figure 2, we plotted the distribution and the profile of mean of drinks per occasion as well as the mediator, normative perception, by group defined by alcohol component intervention. The distribution plots in the bottom of Figure 2 suggest that these outcomes might be Poisson distributed without excessive zeroes.

\begin{figure}
\centering
\includegraphics[width=4in]{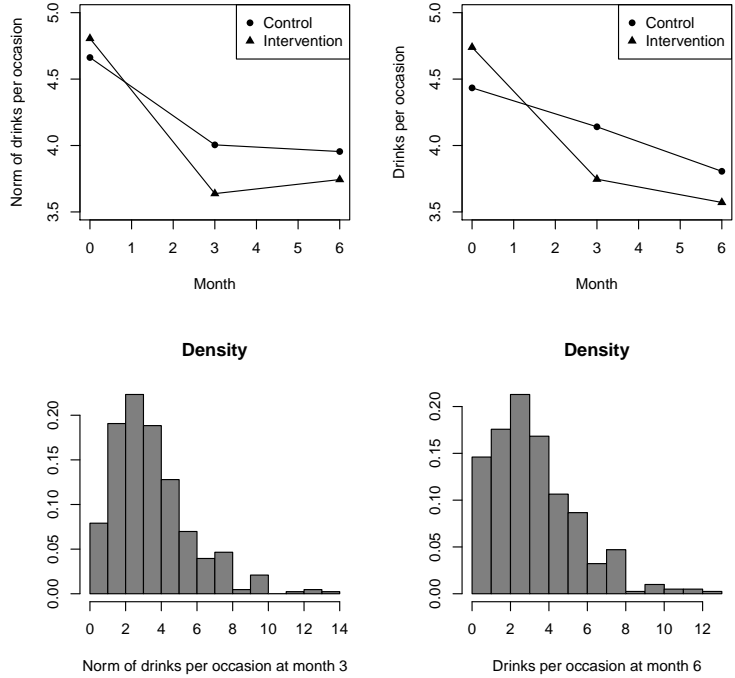}
\caption{Mean and distribution of drinks per occasion and its normative perception, by intervention group}
\end{figure}


\section{Method}
\label{sec:Me}
For the $i$th participant, we denote $Y_{i}$ as the observed outcome, $Z_{i}$ as the intervention, $\bmX_{i}$ as a vector of covariates, and $M_{i}$ as the mediator. The potential outcome $Y_{i}^{zm}$ is defined as the outcome variable that would be observed for $i$th participant if the participant is randomized to intervention level $z$ and receives mediator at level $m$. In graphical model representation, we assume that $\bmU$ is a vector of unmeasured confounders between $M$ and $Y$. Using the notation by Pearl \cite{Pearl_2001}, we can represent the potential outcome as $Y_{i}^{zm}=Y(do$ $Z=z, do$ $M=m, \bmU_{i}, \bmX_{i})$, which means that $Y_{i}^{zm}$ is the outcome if we can manipulate $Z$ and $M$ at level $z$ and $m$. To separate the potential level and observed value, we use lower case to denote potential level while use upper case for random variables observed. We propose the following multiplicative structural mean model:


\begin{equation}
\log E(Y_{i}^{zm}|\bmX_{i},\bmU_i)=g(\bmX_{i},\bmU_i)+\sum_{k=1}^{K}\theta_{k}h_{k}(z,m,\bmX_{i}),
\end{equation}
where the effects of intervention and mediator on the outcome, $h_{k}(\cdot),k=1,2,\cdots,K$, are known
functions, which satisfy $h(0,0,\bmX)=0$ and $g(\cdot)$ is an unknown function representing $\log E(Y_{i}^{00}|\bmX_{i},\bmU_i)$. This model can be used when we assume $Y_i^{zm}|\bmX_{i},\bmU_i$ follows either a Poisson or a negative binomial distribution. Here since $\bmU$ is unobserved, the over-dispersed Poisson model can also be written as a Poisson model by considering an additional unmeasured normally distributed $U$. Compared with the traditional approach to mediation analysis, we allow the unmeasured confounder $\bmU$ in the model with a flexible form. The only requirement is that $U$ does not serve as an effect modifier of both direct and mediator effects. Compared with Imai's nonparametric model, we have a partly linear assumption for the effect of $Z$ and $M$ but Imai \textit{et al.} do not allow the existence of $U$.

Under the stable unit treatment value (SUTVA) and consistency assumptions, we derive the equation on the observed variables as follows:
\begin{equation}
E(Y_{i}|\bmX_{i},Z_{i},\bmM_{i})=G(Z_i,\bmM_i,\bmX_{i})\exp\{\sum_{k=1}^{K}\theta_{k}h_{k}(Z_{i},M_{i},\bmX_{i})\},
\end{equation}
where $G(Z_i,\bmM_i,\bmX_{i})=E[\exp\{g(\bmX_{i},\bmU_i)\}|Z_i,\bmM_i,\bmX_{i}]$. The consistency assumption is a commonly made assumption in causal inference to relate the potential outcome to the observed outcome. The SUTVA assumption means that there is not multiple versions of the treatment, which is needed to clearly define the causal effect, and it also requires no interference.  In the intervention study introduced earlier, the intervention was provided to individuals via the web, and thus, the SUTVA assumption seems reasonable.

Under ignorability of the treatment assumption, which automatically holds for a randomized trial, we have $\bmU_i\perp Z_i|\bmX_i$, which suggests that
\begin{equation}
E(G(Z_i,\bmM_i,\bmX_{i})|\bmX_{i},Z_i)=E(G(Z_i,\bmM_i,\bmX_{i})|\bmX_{i}).
\end{equation}

This implies that for an arbitrary function $A(Z,\bmX)$, the following estimating equation is unbiased:
\beaa
0&=&\sum_iS_i(\bmtheta) \nonumber \\
&=&\sum_i\{A(Z_i,\bmX_i)-E(A(Z_i,\bmX_i)|\bmX_i)\}Y_{i}\exp\{-\sum_{k=1}^{K}\theta_{k}h_{k}(Z_{i},M_{i},\bmX_{i})\}.
\eeaa
Since we have $K$ unknown parameters, we need at least $K$ estimating equations that are not collinear. Computing the expected derivatives of the estimating equations with respect to $\theta$, we obtain that $E\nabla S_i$ has the following expression:
\bea
E\left[\{A(Z_i,\bmX_i)-E(A(Z_i,\bmX_i)|\bmX_i)\}E(G(Z_i,M_i,\bmX_{i})H(Z_i,M_i,\bmX_i)^T|Z,\bmX)\right],
\eea
where $H(Z,\bmM,\bmX)=(h_1(Z,\bmM,\bmX),\cdots,h_K(Z,\bmM,\bmX))^T$. So we need
\beaa
E(Cov H(Z,\bmM,\bmX)|\bmX)>0,
\eeaa
where $>$ means positive definite. The interpretation of this assumption will depend on the specific model. For the simplest model where $K=2$, $h_1(Z_{i},M_{i},\bmX_{i})=Z_i$ and $h_2(Z_{i},M_{i},\bmX_{i})=M_i$, then the assumption above indicates that $\bmX$ modifies the effect of the intervention on mediator. From the view of instrumental variables, the interaction between intervention and $\bmX$ serves as an instrument. We can choose $K$ estimating equations. We can also use the generalized method of moment estimator when the model is over-identified, i.e. we can construct $L>K$ estimating equations that are noncollinear. Then we can use the least square estimator by minimizing
\bea
\|\sum_i\{A(Z_i,\bmX_i)-E(A(Z_i,\bmX_i)|\bmX_i)\}Y_{i}\exp\{-\sum_{k=1}^{K}\theta_{k}h_{k}(Z_{i},M_{i},\bmX_{i})\}\|_{L_2}.
\eea

\subsection{Asymptotic Theorems}
\label{subsec:Asy}
We denote the true value of parameters $\bmtheta$ by $\bmtheta_0$. Under certain regularity conditions, we have
\beaa
\sqrt{n}(\hat{\bmtheta}-\bmtheta_0)\longrightarrow_d N(\bmzero,V),
\eeaa
where
\beaa
V=E(\frac{\partial S_i(\bmtheta)}{\partial \bmtheta})^{-1}E(S_i(\bmtheta)S_i(\bmtheta)^T)E(\frac{\partial S_i(\bmtheta)}{\partial \bmtheta})^{-T}.
\eeaa
Here the expectations are taken under $\bmtheta_0$. This result comes directly from the theory of generalized estimating equations.

The variance-covariance matrix of the estimators, $\hat{\bmtheta}$, can be estimated by a sandwich estimator given by
\beaa
\hat{V}=E_n(\frac{\partial S(\bmtheta)}{\partial \bmtheta})^{-1}E_n(S(\bmtheta)S(\bmtheta)^T)E_n(\frac{\partial S(\bmtheta)}{\partial \bmtheta})^{-T},
\eeaa
where $E_n$ denotes the empirical expectation. Since the empirical expectation will converge to the true expectation uniformly under certain regularity conditions, the estimator $\hat{\bmtheta}$ is consistent. By the uniform law of large numbers, the sandwich estimator is a consistent estimator for the variance covariance matrix.

\subsection{Efficiency Consideration}
Although we do not need to model the covariate effect to obtain a consistent estimator, using a working model for covariate effect can gain efficiency in most cases. After including covariate parts, the estimating can be written as
\bea
0&=&\sum_i\{A(Z_i,\bmX_i)-E[A(Z_i,\bmX_i)|\bmX_i]\}[Y_{i}\exp\{-\sum_{k=1}^{K}\theta_{k}h_{k}(Z_{i},M_{i},\bmX_{i})\}-g(\bmX_i,\bmbeta)].
\eea
Here $g(\bmX_i,\bmbeta)$ is the working model and a simple choice is $g(\bmX_i,\bmbeta)=\exp\{\bmX_i\bmbeta\}$.
To yield an estimator of $\bmbeta$, we can add the following estimating equation:
\beaa
0&=&\sum_i\bmX_i[Y_{i}\exp\{-\sum_{k=1}^{K}\theta_{k}h_{k}(Z_{i},M_{i},\bmX_{i})\}-g(\bmX_i,\bmbeta)].
\eeaa

\section{Simulation}
In this section, we present simulations to show the finite sample performance of our proposed estimator and compare it to the traditional regression estimators. We compared our estimator with the traditional regression method under settings in which sequential ignorability holds or does not hold. The model we considered was a binary intervention $Z$ and a continuous mediator $M$ following the model,
\beaa
E(M|Z,X,U)=\gamma_{z}Z+\gamma_{x}X+\gamma_{zx}ZX+\gamma_{u}U,
\eeaa
and the outcome followed a Poisson model with rate $\exp(\theta_{z}z+\theta_{m}+\theta_uU+\theta_xX)$, or an over-dispersed Poisson model with additional term $\exp(\theta_{z}z+\theta_{m}+\theta_uU+\theta_xX+\theta_vV)$ where $V$ was standard normally distributed and was independent of all other variables. In addition, we simulated the negative binomial model with mean $\exp(\theta_{z}z+\theta_{m}+\theta_uU+\theta_xX)$ and dispersion parameter equal to 2. Here $X$ was standard normally distributed, and $U$ was an independent, standard normally distributed confounder. A non-zero coefficient $\theta_u$ leads to the failure of sequential ignorability assumption. We set the variance for the residual of $M$ as 0.1. We set other parameters as $\theta_x=0.2$, $\theta_{z}=0.1$, $\theta_{m}=0.5$, $\theta_{u}=0,-1$, $\gamma_{z}=0$, $\gamma_{x}=0$, $\gamma_{zx}=1$, $\gamma_{u}=0.5$.

In the simulation, we chose the sample size to be 400 and ran 1000 simulations.  For the traditional regression model for mediation, we used either Poisson regression or negative binomial regression. The simulation results are shown in Figure 3 and Figure 4.

\begin{figure}
\centering
\includegraphics[width=4in]{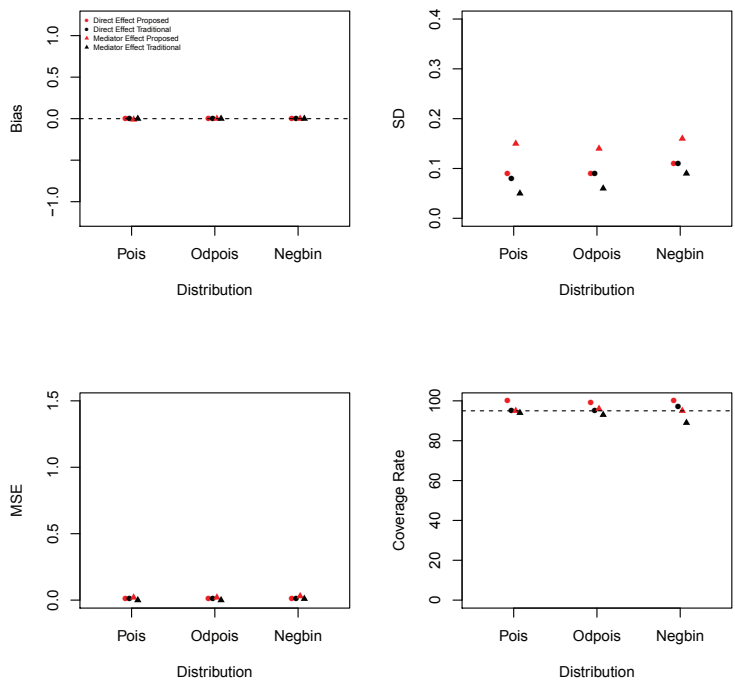}
\caption{Simulation Results without sequential ignorability: Pois means poisson regression, Odpois means over dispersion poisson regression and Negbin means negative binomial regression.}
\end{figure}

\begin{figure}
\centering
\includegraphics[width=4in]{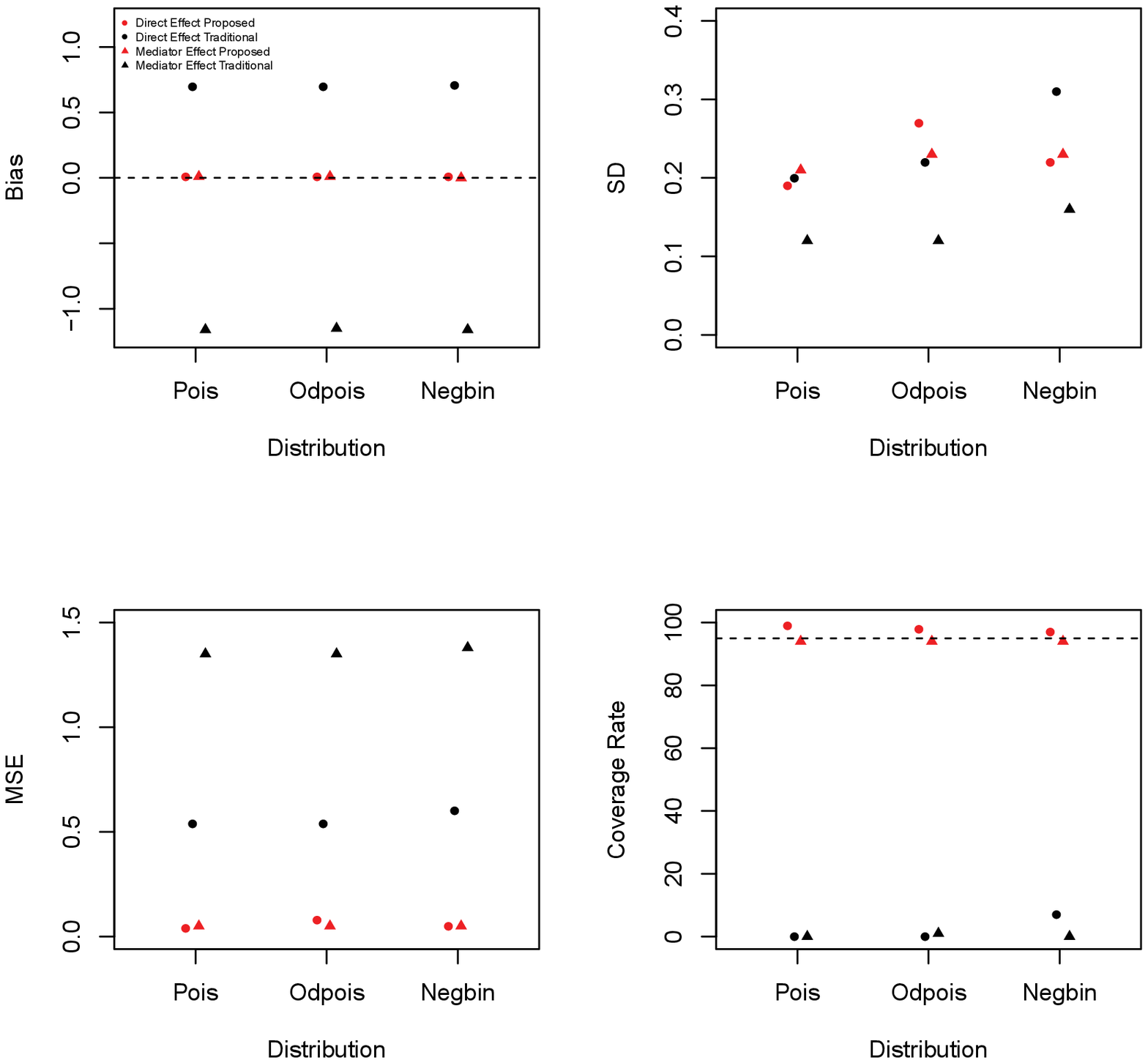}
\caption{Simulation Results with sequential ignorability: Pois means poisson regression, Odpois means over dispersion poisson regression and Negbin means negative binomial regression.}
\end{figure}

Simulation results showed that the proposed method has certain efficiency loss when the sequential ignorability assumption holds. But when the sequential ignorability assumption fails, the traditional regression method can yield highly biased results while the proposed estimator is unbiased. The proposed method has coverage rate higher than its nominal 95\% level and suggests that inference is conservative.

\section{Application to Alcohol Study}
We applied our method to the alcohol use data and compared our estimator with the estimator from traditional mediation models for both drinking frequency and drinks per occasion. To simplify the analyses, we created a binary treatment indicator based on whether the original treatment conditions contained alcohol content or not. So we combined the Alcohol only PNF (n=119), and Combined alcohol and alcohol-related RSB PNF (n=119) groups as Intervention group, and we combined alcohol-related RSB only PNF (n=121) and Control (n=121) groups as Control group. Our goal is to assess whether the alcohol component of the intervention works through changing alcohol-related normative perceptions. We use the 3 months measurement of normative perception as mediator and use the 6 months drinks per occasion as outcome. Among the baseline covariates that are available, we find that a baseline measure of the frequency of casual sex partners serve best as the covariate to construct estimating equation on drinks per occasion, because they have strong interaction with the intervention on normative perceptions. Beyond this variable, we also adjust normative perceptions and drinks per occasion on baseline and gender to increase the efficiency of our proposed estimators. To compare with the traditional method, we run quasi-poisson regression with same set of covariates adjusted. The model is based on the potential outcome and we assumed it has the following form for this example:
\bea
\log E(Y^{zm}|\bmX,\bmU)=\theta_{1}z+\theta_{2}m+g(\bmX,\bmU),
\eea
where $Y^{zm}$ is the 6 months drinks per occasion if the person received treatment z (0: not contain alcohol content; 1: contain alcohol content) and has 3 months normative perception at level m. We did not put any parametric assumption on the covariate effects and the covariates $\bmX$ represents baseline normative perceptions, baseline drinks per occasion and gender. For traditional regression model, we use a parametric model for the covariate effect and the model is based on the observed outcome and we assume it has the following form:
\bea
\log E(Y|\bmX,\bmZ,\bmM)=\theta_{1}\bmZ+\theta_{2}\bmM+\theta_{X}\bmX.
\eea

The results are shown in Table 1. Since the sample size is small and the simulation shows potential over coverage for the asymptotic confidence interval, 500 bootstrap resamples were used to compute standard errors of the proposed estimators.

We find that both our proposed method and the traditional regression method show that 3 months normative perceptions (i.e., mediator) is significantly related to 6 months drinking behavior (OR$>$1), whereas the direct effect is not statistically significant. The estimated mediator effect is stronger using the proposed method compared with the traditional method. Considering that our proposed method is robust to the unmeasured confounder while the traditional regression is not, this result suggest that there are some potential confounders hide the effect of one's normative perceptions on one's drinking behavior. For the traditional regression model, using a negative binomial regression give similar results. Since our model is on the potential outcome, we cannot direct test whether the baseline frequency of casual sex partners modified the direct or mediator effect. But a regression analysis of the observed outcome on intervention, exposure and covariates as well as their interaction does not suggest that there is strong effect modification by baseline frequency of casual sex partners or its norm at baseline. Although the scale of the effect is difference, the sign of the effects and the clinical implications are consistent for both methods. So we confirmed that decreased normative perceptions led to decreased drinking behavior.


\begin{table}[!h]
\begin{center}
\begin{tabular}{|c|c|ccc|}
\hline
Effect &Method & RR &\multicolumn{2}{c|}{95\% CI}\\
\hline
Direct Effect ($\theta_1$)  &Proposed &0.91&0.69&1.20\\
Mediator Effect ($\theta_2$) &Proposed &1.58&1.12&2.22\\
Direct Effect ($\theta_1$) &Traditional&0.94&0.84&1.05 \\
Mediator Effect ($\theta_2$) &Traditional &1.12&1.08&1.16 \\
\hline
\end{tabular}
\end{center}
\tabcolsep 5mm \caption{Results for mediation analysis of 3 month normative perception of drinkes per occasion on 6 month drinks per occasion using both proposed method and Poisson regression method.}
\end{table}

\section{Discussion}
In conclusion, we proposed an alternative method to estimate controlled mediator and direct effects when we are suspicious about the sequential ignorability assumption. Our proposed method can estimate the causal parameters consistently, but loses efficiency when the sequential ignorability holds or when there is a strong risk factor not included in the model. Our analysis yield results that are generally consistent with the original study. This suggests that normative perceptions are causally related to drinking behavior and the association detected in previous study is not due to unmeasured or unidentified confounders. Also, the results suggest that the mediator effect has been slightly under-reported using traditional method.

One extension for our current method is that if we have longitudinal data with more than 3 time points or if we can assume temporal order for the variables measured at same time points. We can assume the models for different time points share some parameters and model them together with a generalized estimating equation. 

As shown in the general model form, theoretically, we can allow multi-arm intervention and do not need to combine the 4 groups into 2. However, it will need stronger interactions between baseline covariates and interventions on the mediator to make sure the weighting functions are far from collinearity. Large sample size is required to obtain any significant results. Given the limited sample size in our example study, we unfortunately cannot obtain much useful information from a 4 group analysis.

Also, we did not model the loss of follow-up procedure in our analysis and use complete data only for the analysis. If we have a model for that procedure, we can use the inverse probability weighting method to handle those missing values, but such method will be sensitive to the correctness of the missing mechanism.

\section*{Acknowledgement}
The data collection is supported by Grant K01 AA016966/National Institute on Alcohol Abuse and Alcoholism.



  \bibliographystyle{elsarticle-harv} 
  \bibliography{mybib}





\end{document}